\documentclass[aps,pra,reprint,groupedaddress,showpacs]{revtex4-1}
\usepackage{mathtools}
\usepackage{amsmath}
\usepackage{multirow}
\usepackage{graphicx}
\usepackage{dcolumn}
\usepackage{bm}
\usepackage{latexsym,epsfig}
\begin{document}


\title{Cluster Mean Field plus Density Matrix Renormalization theory for the Bose Hubbard Model}


\author{$^1$Pallavi P. Gaude, $^2$Ananya Das and $^1$Ramesh V. Pai}
\affiliation{$^1$School of Physical and Applied Sciences, Goa University, Taleigao Plateau, Goa 403 206, India}
\affiliation{$^2$Department of Physics, Parvatibai Chowgule College of Arts and Science-Autonomous, Gogol, Margao, Goa 403 602, India}
\date{\today}

\begin{abstract}
We develop a novel approach to understand the phases of one-dimensional Bose-Hubbard models. We integrate the simplicity of the mean-field theory and the numerical power of the density matrix renormalization group method to build an effective numerical technique with moderate computational resources to determine superfluid order parameters and correlation functions of large one-dimensional systems. We demonstrate the applicability of this method to directly identify superfluid, Mott insulator, and density wave phases in Bose-Hubbard models. 
\end{abstract}


\maketitle


\section{Introduction}
\label{intro}
Experimental advances in ultra-cold atoms in optical lattices have considerable thrust in the study of a many-body quantum system~\cite{bloch1,lewen}.  The prominent example of quantum phase transition is the superfluid (SF) to Mott insulator (MI), and it has been experimentally realized in one-dimension~\cite{stoferle,fertig,mun,haller}, two-dimensions~\cite{spielman1,spielman2,gemelke,bakr}, and 3D~\cite{mun,greiner1,trotzky} optical lattices.  The Bose-Hubbard model, which describes interacting bosons on optical lattices, has been widely used to study the SF to MI phase transition even before the experiments on cold atoms and still holds a significant role in the current studies~\cite{fisher,sheshadri,gultzwiller,jack}.

Cold bosonic atoms in an optical lattice in the tight-binding regime are described by the Bose Hubbard model~\cite{jack},
\begin{align}
\label{eq:bh}
\hat{H} &=-t\sum_{j}\left(\hat{a}^\dagger_{j+1}\hat{a}_{j}
+\hat{a}^\dagger_{j}\hat{a}_{j+1} \right)\nonumber\\
&+\frac{U}{2}\sum_{j}\hat{n}_j(\hat{n}_j-1)
-\mu\sum_{j}\hat{n}_j
\end{align}
where bosons hop between nearest neighboring pairs of site  with amplitude $t>0$,  $ \hat{a}^{\dagger}_{j}$ ($ \hat{a}_{j}$) is the boson creation
(annihilation) operator for the site $j$. The second term is the on-site interaction with strength $U>0$. The last term  controls the boson number  for a given chemical potential $\mu$. The ratio between $U/\nu t$, where $\nu$ is the filling factor (the number of bosons per site), controls the ground state of the BH model. When $U/\nu t \lesssim 1 $ superfluid phase is favored at any filling. Increasing $U/\nu t$ for integer filling quantum fluctuations drive the system into the Mott insulator phase.

Bose-Hubbard model is not exactly solvable even in one dimension. Hence, this model has been studied by several approximate and numerical techniques. Mean-field theories~\cite{fisher,sheshadri,gultzwiller}, numerical techniques such as quantum Monte Carlo simulation ~\cite{capogrosso1,capogrosso2} and strong-coupling-expansion (SCE) techniques ~\cite{freericks,teichmann} for 2D and 3D system and density matrix renormalization group (DMRG)~\cite{rvpaiprl,kuhner,ejima}  and time-evolving block
decimation (TEBD)~\cite{vidal,danshita} for one dimension system have been applied to determine the phases and the critical SF to MI transition point $U/\nu t$.  As such, these methods have certain advantages as well as limitations. Focusing on one-dimensional systems, for example, the simplest of all numerical methods is the mean-field theory, which is exact in the limit of infinite dimension~\cite{sheshadri,gultzwiller}. In the mean-field theory, the BH model~(\ref{eq:bh}) is decoupled from the surrounding lattice, into a single site Hamiltonian which is easily diagonalized. The fluctuations are described by a mean-field superfluid parameter $\psi=\langle \hat a \rangle$.  The ground state energy is minimized with respect to $\psi$. The phases are characterized based on superfluid order parameter which is finite in the SF phases and vanishes in the MI phases. The mean-field theories for model~(\ref{eq:bh}) predict the superfluid and the Mott insulator phases correctly. However, mean-field theories are known to overestimate the superfluid phase boundaries~\cite{cmft}. The Density Matrix Renormalization Group (DMRG), on the other hand, is an effective numerical technique with moderate computational resources to determine the ground state energy and the correlation functions of a large one-dimensional systems~\cite{white}. It has been observed that the DMRG method works well when the ground state has a gap in the energy spectrum~\cite{white,dmrgreview}. When DMRG is applied to Bose-Hubbard model, the quantum phases are determined by analyzing the behavior of the gap in the energy spectrum and the correlation functions such as single-particle density matrix $\langle {\hat a}^\dagger_j {\hat a}_{j+r}\rangle$ and
density-density correlation $\langle {\hat n}_j {\hat n}_{j+r}\rangle$~\cite{rvpaiprl,kuhner,ejima}. The DMRG method generally works in the canonical ensemble. Hence the superfluid order parameter $\psi=\langle {\hat a}\rangle=0$ in all phases. 

Several extensions of the Bose-Hubbard model, notably the extended Bose-Hubbard model, spin-1 Bose-Hubbard model, show exotic gapless phases like supersolid, polar/Ferro superfluid, and pair superfluids.  It is desirable to determine the superfluid order parameter to characterize these exotic phases. Unlike the mean-field theories, the DMRG method can't resolve these phases directly due to its limitation in determining superfluid order parameters. 

The cluster mean-field theory (CMFT), which is an extension of the single-site mean-field theory considers a cluster of sites in the build-up of mean-field Hamiltonian~\cite{cmft}. It has been reported that CMFT improves the phase boundary compared to simple single site mean-field theory~\cite{cmft,bhargav}. However, there are limitations in forming more extensive cluster sizes as the Hilbert space of the cluster increases exponentially with the number of sites.
	
In this paper, we provide a new approach,  which utilizes the DMRG capability to handle larger system sizes and the simplicity of the CMFT method.  In this way, this new approach captures the success of both the DMRG and the CMFT methods. The primary aim of this work is to demonstrate this new approach, which we call CMFT+DMRG for the Bose Hubbard Model, and test and compare with the DMRG method.

This paper is organized as follows: Section~\ref{model} describes the CMFT+DMRG formalism. The results and the comparisons are given in Section~\ref{results}. Finally, we conclude our work in Section~\ref{conclusions}.

\section{CMFT+DMRG Method}
\label{model}

First, we set up to solve the model (\ref{eq:bh}) in the cluster mean-field framework~\cite{cmft,bhargav}. The whole lattice is partitioned into $N_C$ clusters with each cluster having $L$ number of sites. The Hamiltonian (\ref{eq:bh}) is then written as
\begin{align}
\hat{H}=\sum_{p}\hat{H}_p^{loc}+\sum_p \hat{H}_p^{hop}
\label{eq:bhc}
\end{align}
where $p$ represents the cluster index and
\begin{align}
\hat{H}_p^{loc} &=-t\sum_{j}\left(\hat{a}^\dagger_{p,j+1}\hat{a}_{p,j}
+\hat{a}^\dagger_{p,j}\hat{a}_{p,j+1} \right)\nonumber\\
&+\frac{U}{2}\sum_{j}\hat{n}_{p,j}(\hat{n}_{p,j}-1)
-\sum_{j}^{}\mu\hat{n}_{p,j}.
\label{eq:bhloc}
\end{align}
Here  $ \hat{a}^{\dagger}_{p,j}$ ($ \hat{a}_{p,j}$) is the boson creation
(annihilation) operator for the site $j$ in the cluster $p$ and $\hat{n}_{p,j}=\hat{a}^\dagger_{p,j}\hat{a}_{p,j}$ is the number operator.
The second term in the Hamiltonian Eq.~(\ref{eq:bhc}) represents the hopping of bosons between the clusters and is given by
\begin{align}
\hat{H}_p^{hop}=
 -t \left(\hat{a}^\dagger_{p,1}\hat{a}_{p-1,L}
+\hat{a}^\dagger_{p-1,L}\hat{a}_{p,1} \right).
\label{eq:bhhop}
\end{align}
We now decouple each cluster  from its neighbour clusters by using standard mean-field decoupling i.e., $a_{p,j}=\langle a_{p,j}\rangle+\delta a_{p,j}$ where $\langle a_{p,j} \rangle=\psi_{p,j}$ is the superfluid order parameter.
Considering the fluctuation $\delta a_{p,j}$ to be small and thus neglecting second-order fluctuations, we approximate,
\begin{align}
\hat{a}^\dagger_{p,1}\hat{a}_{p-1,L}
+\hat{a}^\dagger_{p-1,L}\hat{a}_{p,1} & \approx
\hat{a}^\dagger_{p,1} \psi_{p-1,L}+\hat{a}_{p,1}\psi^*_{p-1,L} \nonumber \\
&-\frac{1}{2}\left( \psi^*_{p-1,L} \psi_{p,1}+\psi_{p-1,L} \psi^*_{p,1}\right)\nonumber \\ & +\hat{a}^\dagger_{p,L}\psi_{p+1,1}+\hat{a}_{p,L}\psi^*_{p+1,1}\nonumber \\
&-\frac{1}{2}\left( \psi^*_{p+1,1} \psi_{p,L}+\psi_{p+1,1} \psi^*_{p,L}\right).
\end{align}
Assuming, without loss of generality, the superfluid order parameter $\psi_{p,j}$ to be real and homogeneous, Eq.~(\ref{eq:bhhop}) is written as
\begin{align}
\hat{H}_p^{hop}=
&-t \left(
(\hat{a}^\dagger_{p,1}+\hat{a}_{p,1})\psi-|\psi|^2 \right) \nonumber \\
&-t\left((\hat{a}^\dagger_{p,L}+\hat{a}_{p,L})\psi-|\psi|^2 \right)
\label{eq:bhhop1}.
\end{align}
Using Eqs.~(\ref{eq:bhloc}) and (\ref{eq:bhhop1}) in Eq.~(\ref{eq:bhc}), we get
\begin{align}
\hat{H}=\sum_{p}\hat{H}_p^{C}
\label{eq:bhc1}
\end{align}
where $\hat{H}_p^{C}$ is the Hamiltonian for a cluster of $L$ sites. Dropping the cluster index $p$
\begin{align}
\hat{H}^{C}= &=-t\sum_{j=1}^{L-1}\left(\hat{a}^\dagger_{j+1}\hat{a}_{j}
+\hat{a}^\dagger_{j}\hat{a}_{j+1} \right)\nonumber\\
&+\frac{U}{2}\sum_{j=1}^L\hat{n}_{j}(\hat{n}_{j}-1)
-\sum_{j=1}^{L}\mu\hat{n}_{j}\nonumber \\
&-t \left(
(\hat{a}^\dagger_{1}+\hat{a}_{1})\psi-|\psi|^2 \right) \nonumber \\
&-t\left((\hat{a}^\dagger_{L}+\hat{a}_{L})\psi-|\psi|^2 \right).
\label{eq:bhc2}
\end{align}

This cluster Hamiltonian has been studied in different limits. For example, in the limit, $L=1$ Eq.~\ref{eq:bhc2} is nothing but the   single site mean-field theory Hamiltonian~\cite{sheshadri,gultzwiller}. McIntosh et al. have considered this Hamiltonian with cluster size up to $L=8$~\cite{cmft}. DMRG method exploits the Bose-Hubbard Hamiltonian in the canonical ensemble with a fixed number of particles. If we neglect the last three terms in the Hamiltonian (\ref{eq:bhc2}), we get the Bose-Hubbard model in the canonical ensemble and has been studied using DMRG to obtain an accurate phase diagram~\cite{rvpaiprl,kuhner,ejima}.

Because the DMRG method works in the canonical ensemble the number of particles is fixed, hence the superfluid order parameter $\psi=\langle a \rangle=0$ in all phases. However, the cluster Hamiltonian works in the grand-canonical ensemble and commutation $[\hat{H}^C,\hat{N}]\ne 0$. The superfluid parameter  $\psi=\langle a \rangle$ can be finite, and such phase can be identified as the superfluid phase. In the Mott insulator phase, $\psi=0$ and the Hamiltonian $\hat{H}^{C}$ commutes with the $\hat{N}$, which implies that our study can reproduce earlier DMRG results. Thus, the CMFT+DMRG method interplay between single-site mean-field theory to the DMRG in the one-dimension Bose Hubbard model.

The task now is to obtain the ground state energy and the wave function of the Hamiltonian (\ref{eq:bhc2}) for any given length $L$ using the CMFT+DMRG method. We describe these steps below.
 \begin{description}
 	\item[ Step I] Consider a lattice of small size $l$, say $l=1$ forming the system block $S$. The Hilbert space of $S$ has dimension $M^S$ and is represented by states $\left\{ |\mu^S_l \rangle \right\}$. (For example, if $l=1$, $\left\{ |\mu^S_l \rangle \right\}$ can be Fock states $\{ |0 \rangle , |1 \rangle , |2 \rangle , \cdots, |n_{max}\rangle$; $n_{max}$ being the maximum number of bosons allowed per site and  $M^S=n_{max}+1$. It may be noted that $n_{max}=\infty$ for bosons, however, for numerical calculation, we will truncate Fock states to $M^S$ states. The value of $n_{max}$ depends on the parameters of the model (\ref{eq:bh}) such as $U/t$ and $\mu$).   Obtain the Hamiltonian ${\hat H}_l^S$ and operators acting on the block. Similarly, form an Environment block $E$.

 	\item[Step II] Form a new system block $S'$ from $S$ and one added site as shown in Fig.~\ref{fig:newsystem}. Hilbert space of the new system block $S'$ has dimension $M^S\times N_{s}$ and is represented by states $\left\{ |\mu^S_l\rangle \mid \sigma^S \rangle \right\}$. Here $N_s=n_{max}+1$ is the number of states per site.  Similarly, form an Environment block $E'$.
 	 \begin{figure}[htb!]
 		\centering
 		\includegraphics[width=6cm, height=1cm]{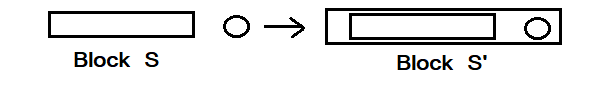}
 		\caption {New system block $S'$ is formed from system block $S$ and one added site represented by open circle.}
 		\label{fig:newsystem}
 	\end{figure}
 	\item[Step III] Now build a superblock of length $L=2l+2$ as shown in Fig.~\ref{fig:superblock}. Construct the Hamiltonian matrix ${\hat H}_{2l+2}$ for a given initial guess for $\psi$ and find the ground state energy $E_L(\psi)$ and the wave function by large sparse-matrix diagonalization. This is the most time-consuming step in this algorithm. Minimize the ground state energy $E_L(\psi)$ with respect to $\psi$ to obtain global ground state energy $E_{GS}$, the wave function $|\Psi_{GS}\rangle$ and the superfluid order parameter $\psi_j=\langle \Psi_{GS}| {\hat a}_j|\Psi_{GS} \rangle$.
 	
 	The ground state wave function is given by
 	\begin{align}
 |\Psi_{GS}\rangle&=\sum_{S'E'} C_{S'E'} |S'~E'\rangle\\ \nonumber &=\sum_{\mu_l^S \sigma^S \sigma^E\mu_l^E}C_{\mu_l^S \sigma^S \sigma^E\mu_l^E}|{\mu_l^S \sigma^S \sigma^E\mu_l^E}\rangle
 	\end{align}

  \begin{figure}[htb!]
 	\centering
 	\includegraphics[width=6cm, height=1cm]{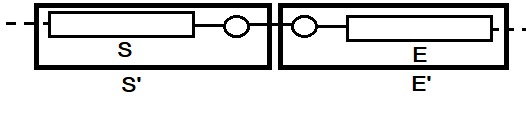}
 	\caption {Super block of length $L=2l+2$. The continuous lines represent hopping term ($-t({\hat a}^\dagger_{j}{\hat a}_{j+1}+H.C))$ in the Eq.~\ref{eq:bhc2} and dashed lines represent $-t(({\hat a}^\dagger_1+{\hat a}_1)\psi -\mid \psi^2\mid)$ and $-t(({\hat a}^\dagger_L+{\hat a}_L)\psi -\mid \psi^2\mid)$.  }
 	\label{fig:superblock}
 \end{figure}

 	\item[Step IV] Construct a reduced density-matrix ${\hat \rho}_{S'}$ for the block system $S'$.
 	\begin{equation}
 {\hat \rho}_{S'}=Tr_{E'}|\Psi_{GS}\rangle \langle \Psi_{GS}\mid
 	\end{equation}
 That is
 	\begin{equation}
 \langle \mu_l^S\sigma^S|{\hat \rho}_{S'}|\nu_l^S\tau^S\rangle=\sum_{\sigma^E\mu_l^E}
 C^*_{\mu_l^S \sigma^S \sigma^E\mu_l^E}C_{\nu_l^S \tau^S \sigma^E\mu_l^E}.
 \end{equation}	
 Diagonalize ${\hat \rho}_{S'}$ to obtain its eigenvectors
 \begin{equation}
 |\alpha\rangle= \sum_{\mu_l^S\sigma^S}O^\alpha_{\mu_l^S\sigma^S}|\mu_l^S\sigma^S\rangle.
 \label{eq:o}
 \end{equation}
 and the eigenvalues $\omega_\alpha$.
 $\omega_\alpha$ measures the weight of the state  $|\alpha\rangle$ in the $|\Psi_{GS}\rangle$ and satisfy  $\sum_{\alpha}\omega_\alpha=1$. Form a new
 (reduced) basis for $S'$ by taking the $M^S$ eigenstates  with the largest weights $\omega_\alpha$.
 The new basis is represented by $M^S$ eigenstates of the reduced density matrix. This way we have truncated the Hilbert basis of the system block $S'$ from $M^S\times N^S$ to $M^S$. This is the most important step of the DMRG method.

 Transform ${\hat H}^{S'}_{l+1}$ and operators to the new basis. i.e.,
 \begin{equation}
{\hat H}^{S'}_{new}=O^\dagger {\hat H}^{S'} O
 \end{equation}
 where $O$ is $M^S\times N^S$ rectangular transformation matrix from Eq.~\ref{eq:o}. Proceed likewise for the environment.

  	\item[Step V] Repeat Step I to IV with block size $l+1$ and continue the iteration until the desired length $L$. The system size is increased by 2 in each iteration.  Calculate the ground state properties (energies, order parameters,
 	and correlators for all $L$).
 	 \end{description}

 \section{Results}
 \label{results}
\subsection{Bose-Hubbard Model}
  \begin{figure}[htb!]
 	\centering
 	\includegraphics[width=8cm, height=8cm]{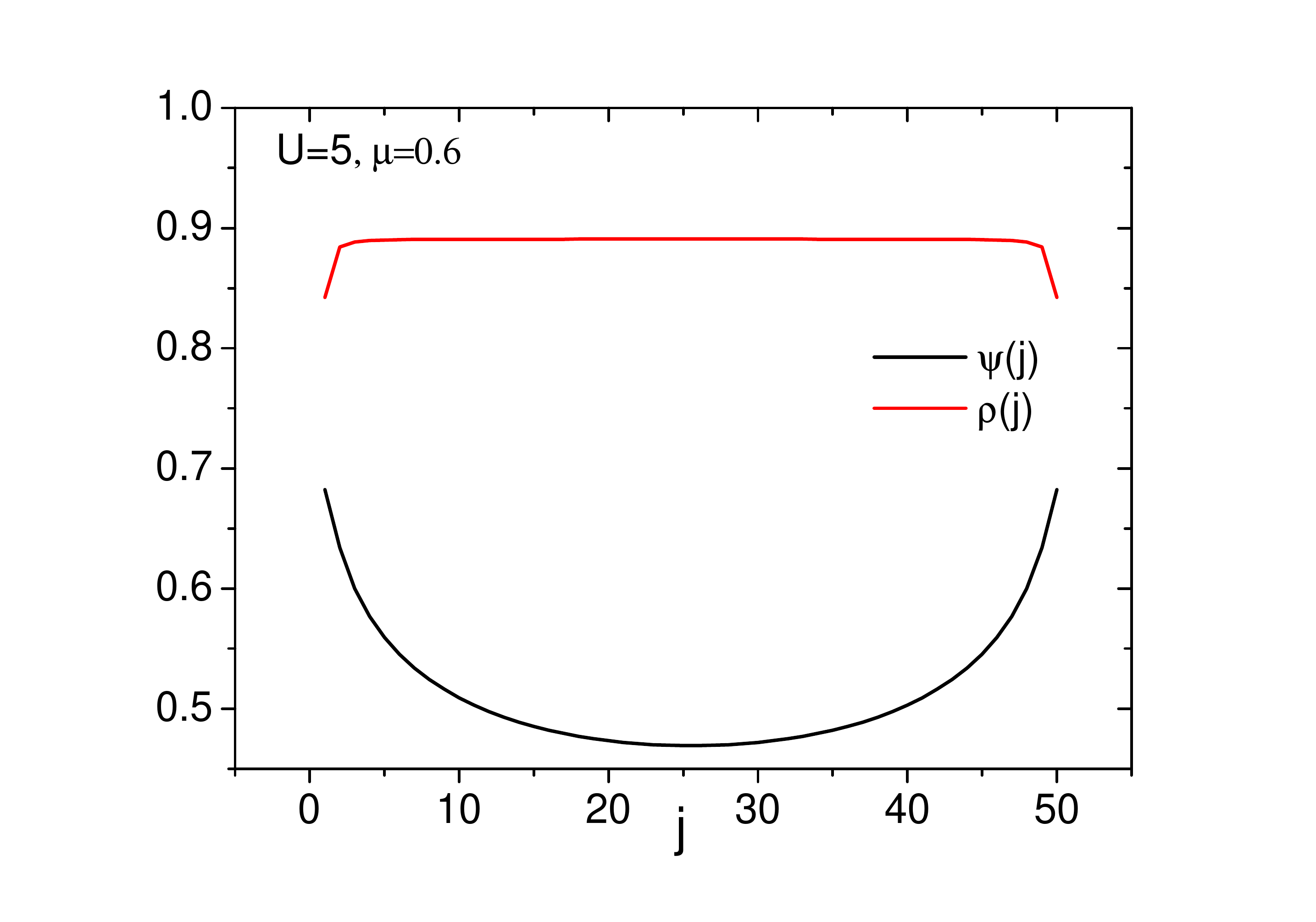}
 	\caption {(Color online) Superfluid order parameter $\psi(j)$ and boson density $\rho(j)$ are plotted as a function of position $j$. Here we have considered a lattice with length $L=50$ and model parameters $U=5$ and $\mu=0.6$.   }
 	\label{fig:fig1}
 \end{figure}

  We now discuss the results of the Bose-Hubbard model. The CMFT+DMRG calculations are performed by retaining $M^S=50$ eigenstates in the left/right block reduced density matrix and taking $n_{max}=3$ which, is found to be sufficient if we restrict the density of bosons $\rho<2$ as done in the present report. The neglected truncated weight
 $\epsilon=1-\sum_{\alpha=1}^{M^S}\omega_\alpha$ is of the order of $10^{-8}$. We set the energy scale by $t=1$.

\begin{figure}[htb!]
	\centering
	\includegraphics[width=8cm, height=8cm]{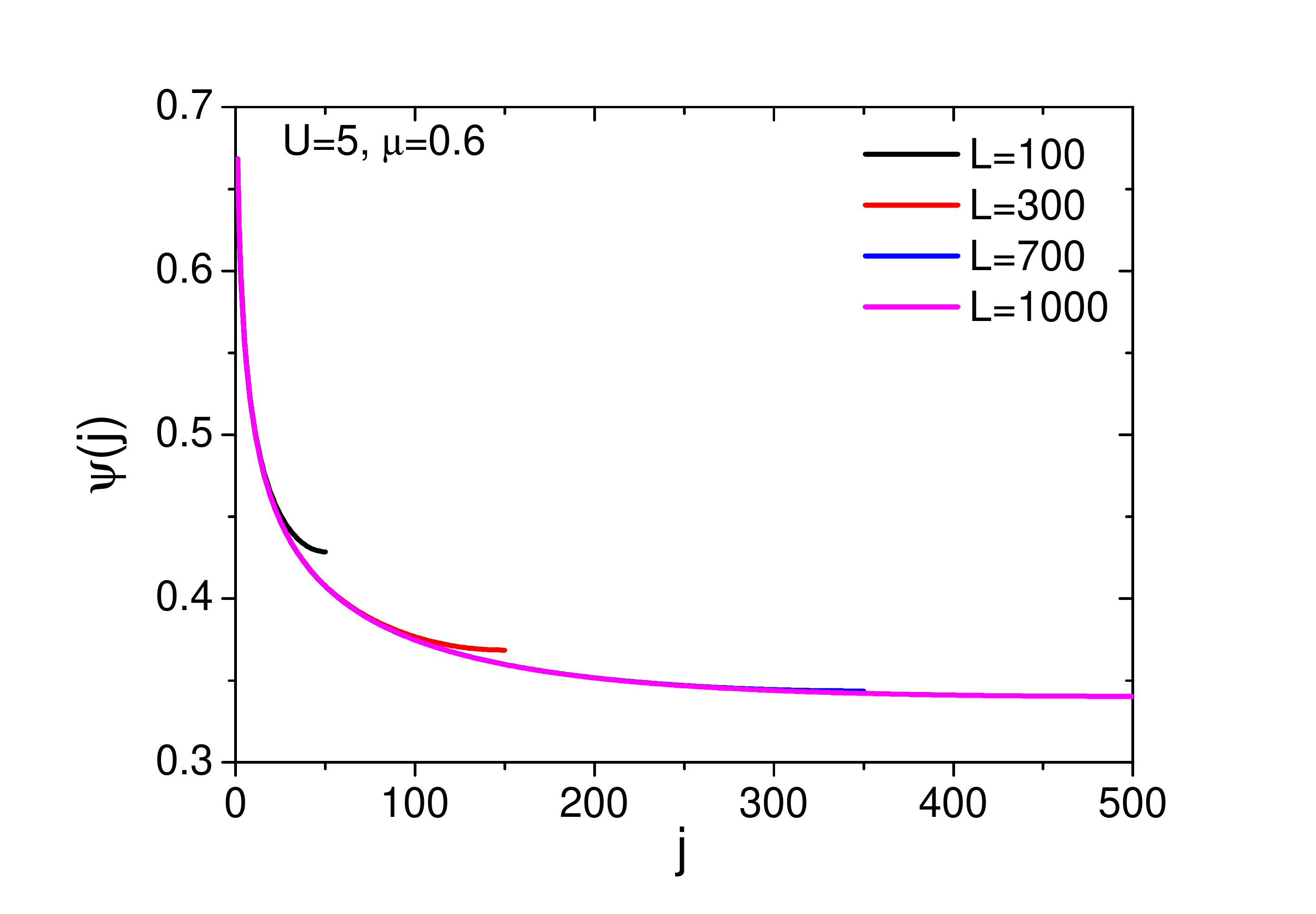}
	\caption {(Color online)The superfluid order parameter $\psi(j)$ are plotted as a function of position $j$ for different lengths $L$. Here $U=5$ and $\mu=0.6$.}
	\label{fig:fig1b}
\end{figure}

\begin{figure}[htb!]
	\centering
	\includegraphics[width=8cm, height=8cm]{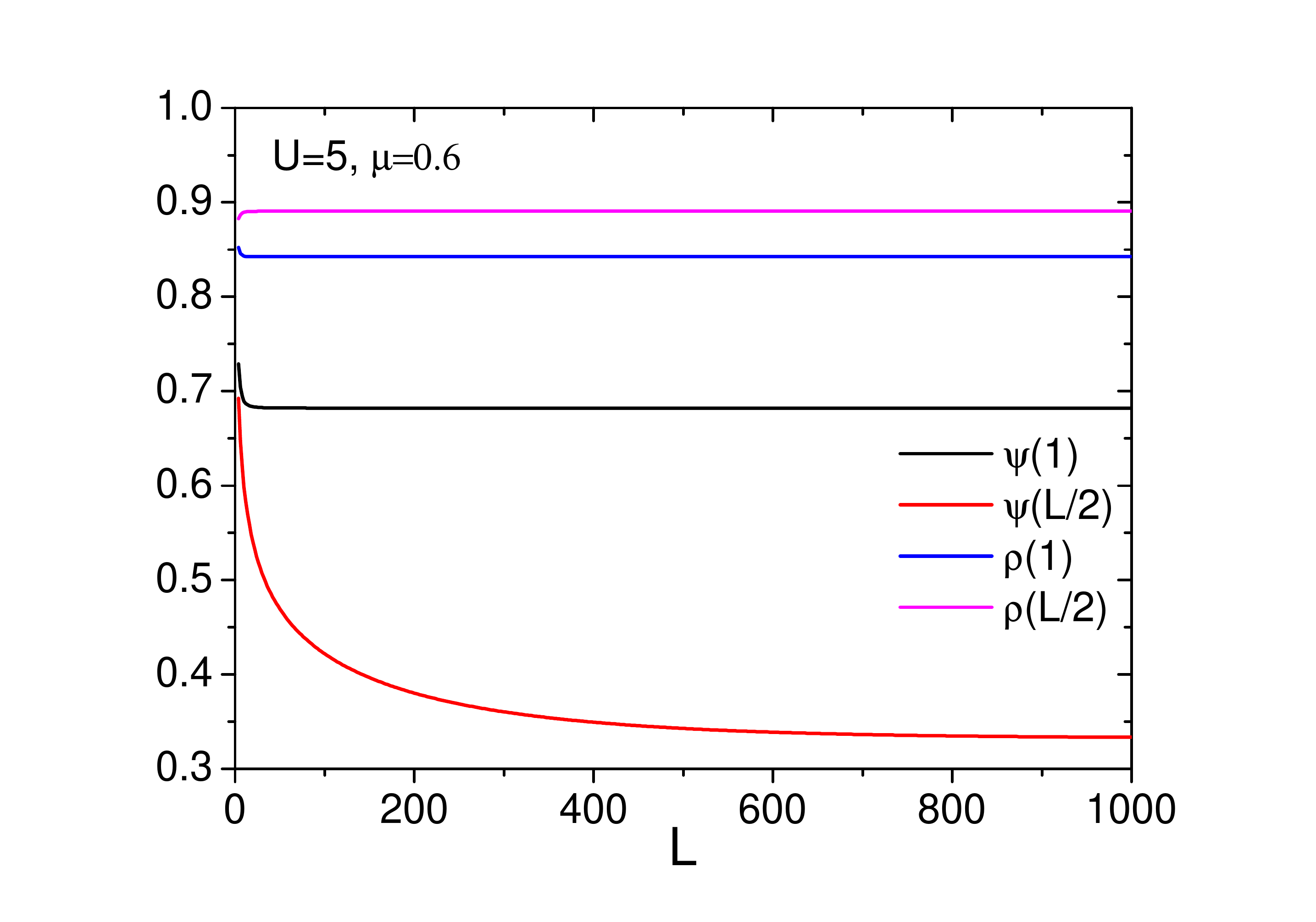}
	\caption {(Color online)The superfluid order parameters $\psi(1)$, $\psi(L/2)$ and the boson densities $\rho(1)$ and $\rho(L/2)$ are plotted as a function length $L$ for $U=5$ and $\mu=0.6$.}
	\label{fig:fig2a}
\end{figure}

\begin{figure}[htb!]
	\centering
	\includegraphics[width=8cm, height=8cm]{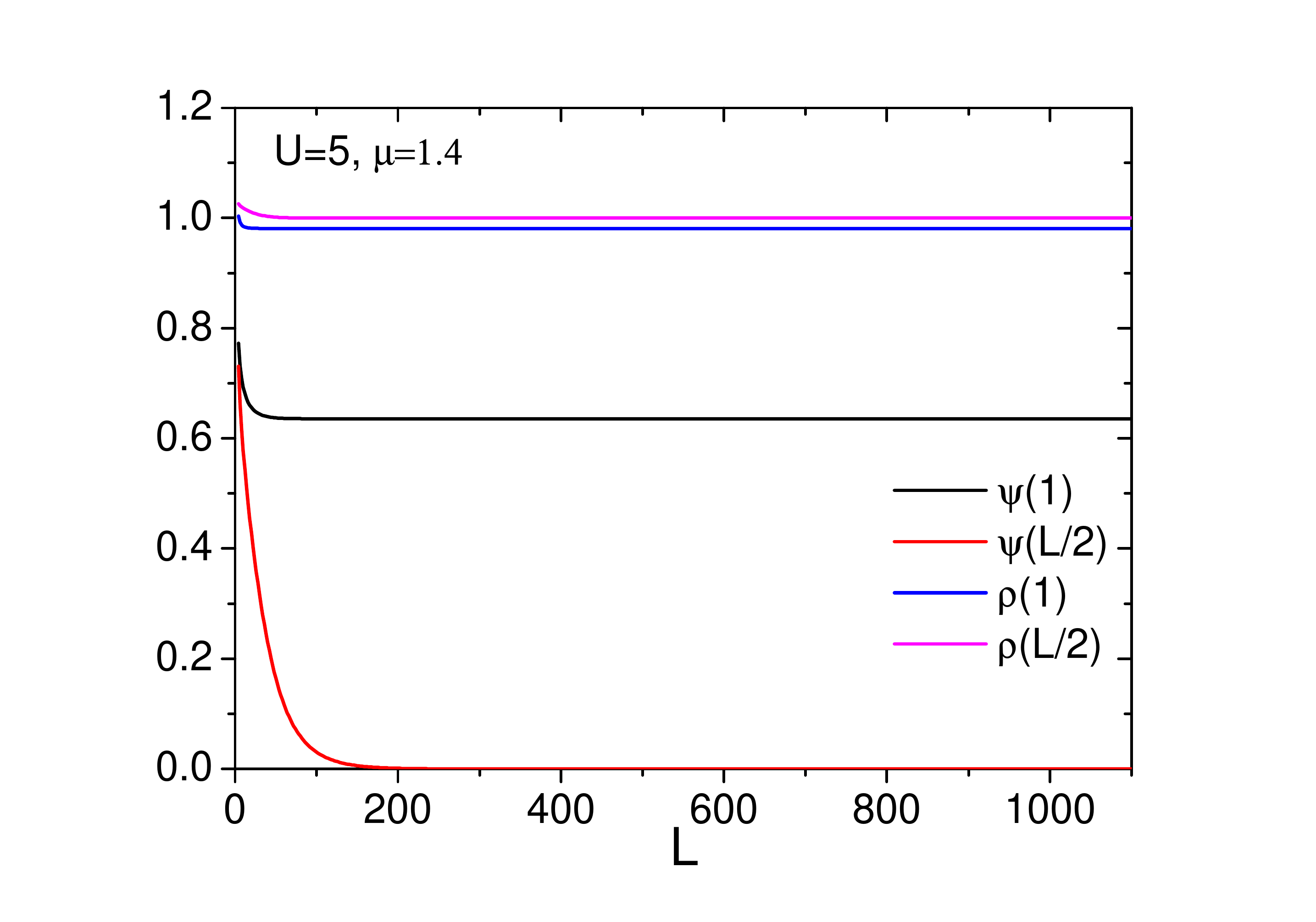}
	\caption {(Color online)The superfluid order parameters $\psi(1)$, $\psi(L/2)$ and the boson densities $\rho(1)$ and $\rho(L/2)$ are plotted as a function length $L$ for $U=5$ and $\mu=1.4$.}
	\label{fig:fig2b}
\end{figure}

  We begin our discussion by analyzing the behavior of the superfluid order parameters $\psi(j)$, and density of bosons $\rho(j)$.  For $U=5$, and $\mu=0.6$, we plot $\psi(j)$ and $\rho(j)$ as a function of lattice position $j$ for  a system length with $L=50$ in Fig.~\ref{fig:fig1}. This depicts a typical behaviour of $\psi(j)$ and $\rho(j)$. The edge sites have higher superfluid order parameter values compared to the center. $\psi(j)$ decreases as the lattice position $j$ moves away from the edges and has the least value when $j=L/2$. Similarly, density $\rho(j)$ increases from the edges to the center. This behavior of $\psi(j)$ is not difficult to understand. The mean-field approximation affects the edge sites and as the lattice position moves away from the edges, the effect of the approximation tampers off. The mean-field approximation is known to overestimate the superfluid phase, hence, the values of the superfluid order parameter are larger at the edges compared to the center.  In order to understand the converges of the superfluid order parameters with system length $L$, we plot $\psi(j)$ for different lengths; $L=100, ~300, ~700, ~\mbox{and,~} 1000$ in Fig.~\ref{fig:fig1b}. We observe that the superfluid order parameters start converging from the edges as the system length $L$ increases.  For example, for $L=100$, $\psi(j)$ have been converged for all $j$ except near the center of the lattice $j\sim 50$. As length increases, $\psi(j)$ converges for more range of values of $j$ and eventually, for larger $L$,   $\psi(j)$ converges for the entire system. To demonstrate this behaviour further, we plot  $\psi(1)$, $\psi(L/2)$, $\rho(1)$ and $\rho(L/2)$ in Figs.~\ref{fig:fig2a} and \ref{fig:fig2b}, respectively, for $\mu=0.6$ and $1.4$ keeping $U=5$.  The densities, $\rho(1)$ and $\rho(L/2)$, and the superfluid parameter for the edge site $\psi(1)$ converge faster with $L$ compare to $\psi(L/2)$. $\psi(L/2)$ converges eventually as length increases further. For $\mu=0.6$, $\psi(L/2)$ converged to a finite value which implies a superfluid phase. However, for $\mu=1.4$, $\psi(L/2)$ converge to zero yielding a Mott insulator phase. It may be noted that $\psi(1)$ is finite for both cases. We conclude from the above  behavior of convergence of superfluid order parameters and densities that $\psi(L/2)$ and $\rho(L/2)$ can be taken as the superfluid fluid order parameter and density of system with length $L$. We denote these by $\psi_L$ and $\rho_L$, respectively.

\begin{figure}[htb!]
	\centering
	\includegraphics[width=8cm, height=8cm]{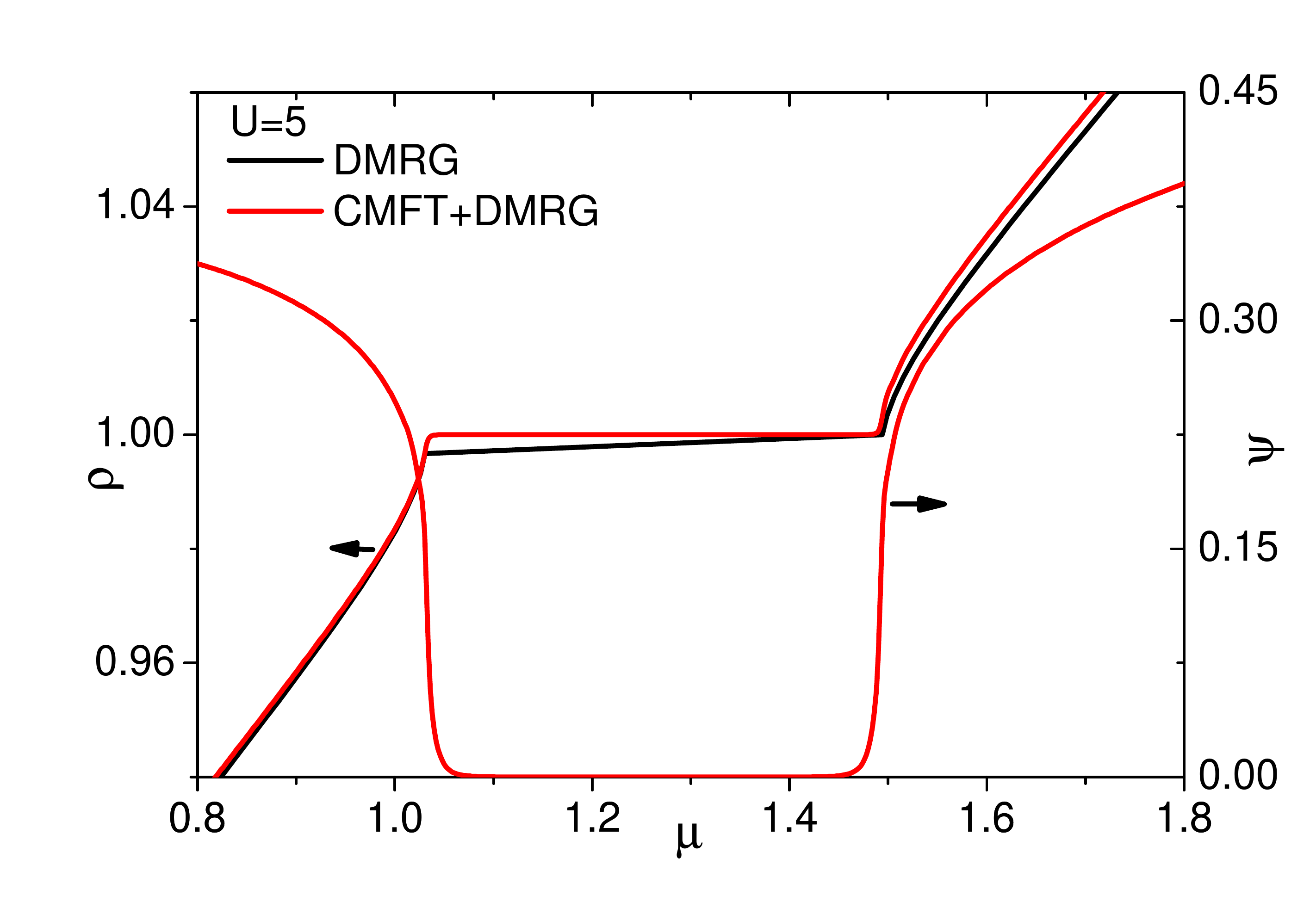}
	\caption {(Color online)Comparison between density of boson $\rho$ obtained with DMRG (black line with bullet) and CMFT+DMRG (red line with bullet) methods. Also plotted superfluid order parameter $\psi$ calculated using CMFT+DMRG method.}
	\label{fig:fig3}
\end{figure}

 We now compare the CMFT+DMRG result with the standard DMRG. In DMRG, the ground state energy $E_L(N)$ of the system of length $L$ having $N$ bosons is obtained using finite-size DMRG procedure~\cite{ramanan}. The density of the system $\rho_L=N/L$ and chemical potential corresponding to this density is determined using the relation
 
\begin{equation}
\mu=(\mu^++\mu^-)/2,
\label{eq:dmrg_mu}
\end{equation}
where $\mu^{\pm}=E_L(N+1)\pm E_L(N)$. We plot the density calculated using the DMRG method and the CMFT+DMRG method for different chemical potentials in Fig.~\ref{fig:fig3} for the system of length $L=300$ and $U=5$. We observe that the density obtained from both methods agree with each other. Density increases with chemical potential and remains pinned at $\rho=1$ for a range of $\mu$ values.  This region corresponding to the Mott insulator phase has finite gap $\Delta=\mu^+-\mu^-$ and vanishing compressibility $\kappa=\left(\frac{d\rho}{d\mu}\right)$. The region outside $\rho=1$  has finite compressibility. This region is considered a superfluid phase.  The DMRG method, unlike the CMFT+DMRG method, however, doesn't have access to superfluid order parameters to identify SF and MI phases directly.  We plot the superfluid order parameter obtained from the CMFT+DMRG method in the same figure. We observe that $\psi$ vanishes in the Mott insulator phase as it should be and is finite in the superfluid phase. Thus the CMFT+DMRG method gives direct access to different phases in the model.

\begin{figure}[htb!]
	\centering
	\includegraphics[width=8cm, height=8cm]{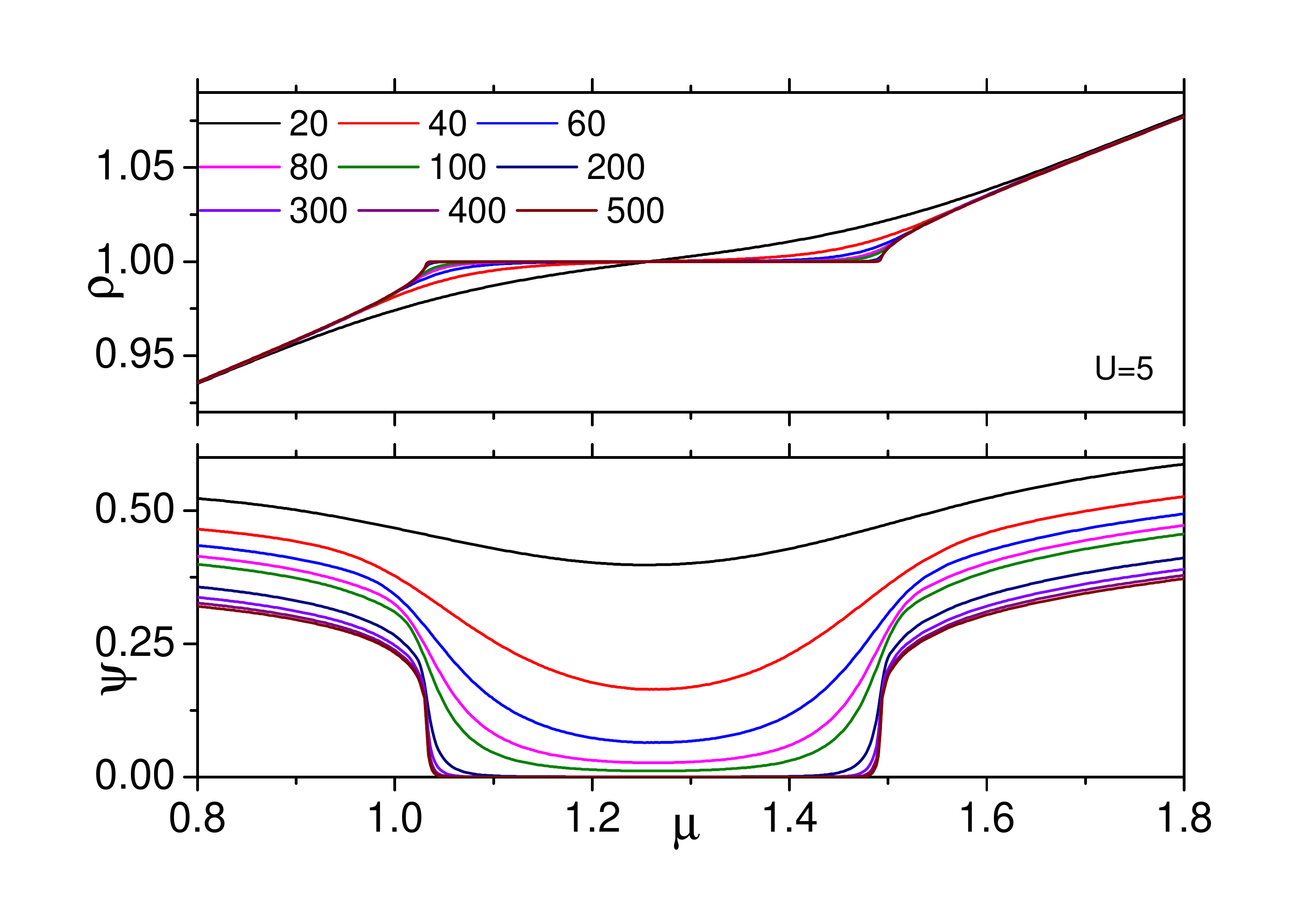}
	\caption {(Color online) (a) SF order parameter $\psi$ and (b) density $\rho$ for different lengths $L=100,~300,~500$, for $U=5$.}
	\label{fig:fig4a}
\end{figure}

\begin{figure}[htb!]
	\centering
	\includegraphics[width=8cm, height=8cm]{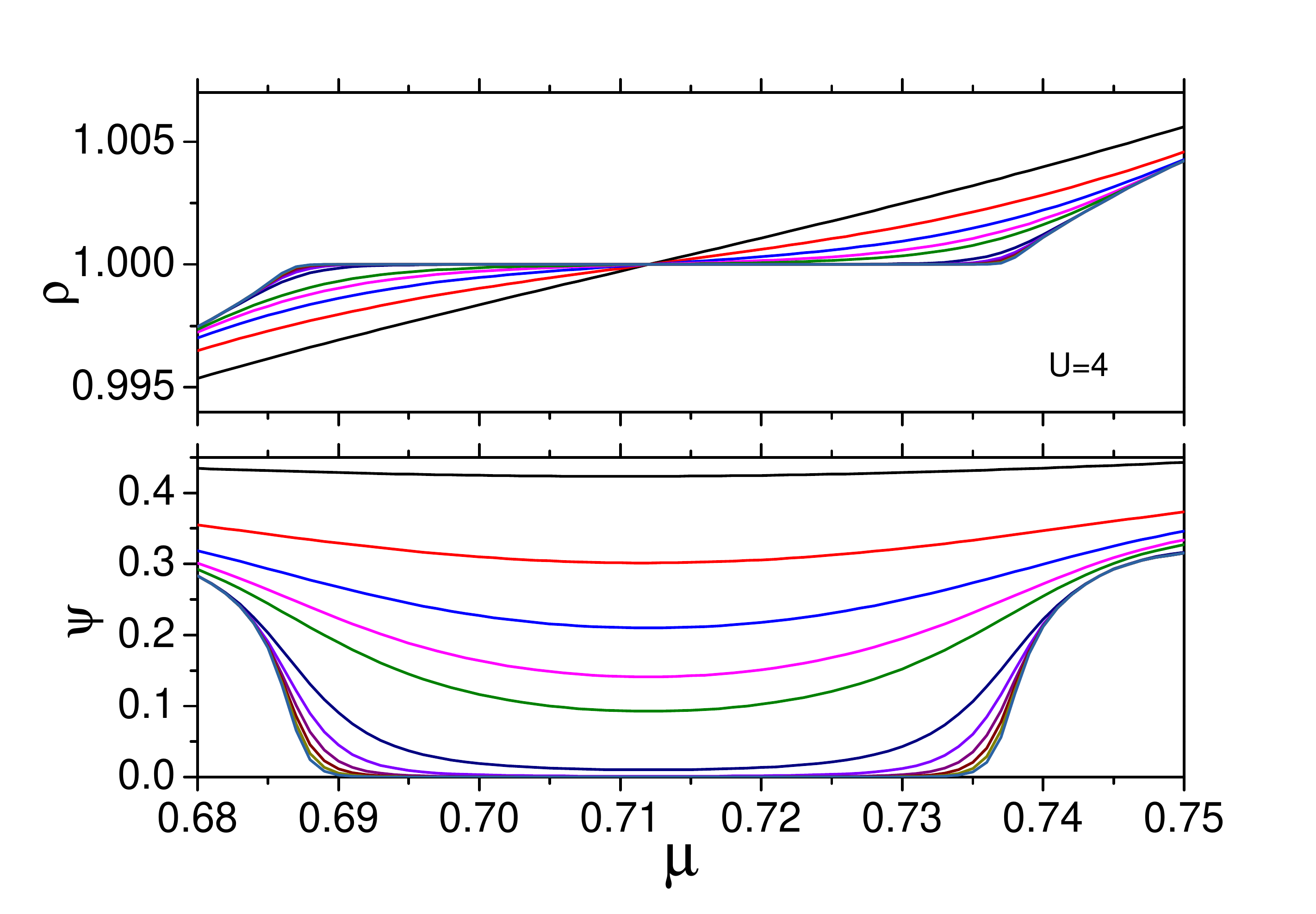}
	\caption {(Color online) (a) SF order parameter $\psi$ and (b) density $\rho$ for different lengths $L=100,~200,~\cdots,~3700$.}
	\label{fig:fig4c}
\end{figure}

 We now address the question of convergence of the superfluid order parameter $\psi$ and density $\rho$ to the system length $L$. The convergence of $\psi$ and $\rho$ depend on (i) the value of the on-site interaction $U$ compared to $U_C$ and (ii) the value of chemical potential $\mu$ compared to $\mu^{-/+}$. Here $U_C$ is the critical on-site interaction for SF-MI transition for $\rho=1$ and $\mu^-(U)$ and $\mu^+(U)$ are the lower and upper edge of the Mott lobe for a given $U$. If $U>>U_C$ and $|\mu-\mu^{-/+}(U)|>> 0$, the SF order parameter and density convergence rapidly with $L$. However, in the opposite limit i.e., $U \sim U_C$ and  $|\mu-\mu^{-/+}(U)| \sim 0$ the convergence is very slow. In these limits, the correlation length $\xi$ is large, and the convergence of the superfluid order parameter is guaranteed if and only if $L>>\xi$.   To demonstrate this behavior, we plot  $\psi_L$ and $\rho_L$ obtained from CMFT+DMRG for $U=5$ and $U=4$ in Fig.~\ref{fig:fig4a} and Fig~\ref{fig:fig4c}, respectively . It may note that the best estimate of $U_C\sim 3.3$. For $U=5$, length $L=500$ is sufficient for the convergence of the SF order parameter. However, as we decrease the on-site interaction, say $U=4$, the convergence is slow and requires a larger length.

\begin{figure}[htb!]
	\centering
	\includegraphics[width=8cm, height=8cm]{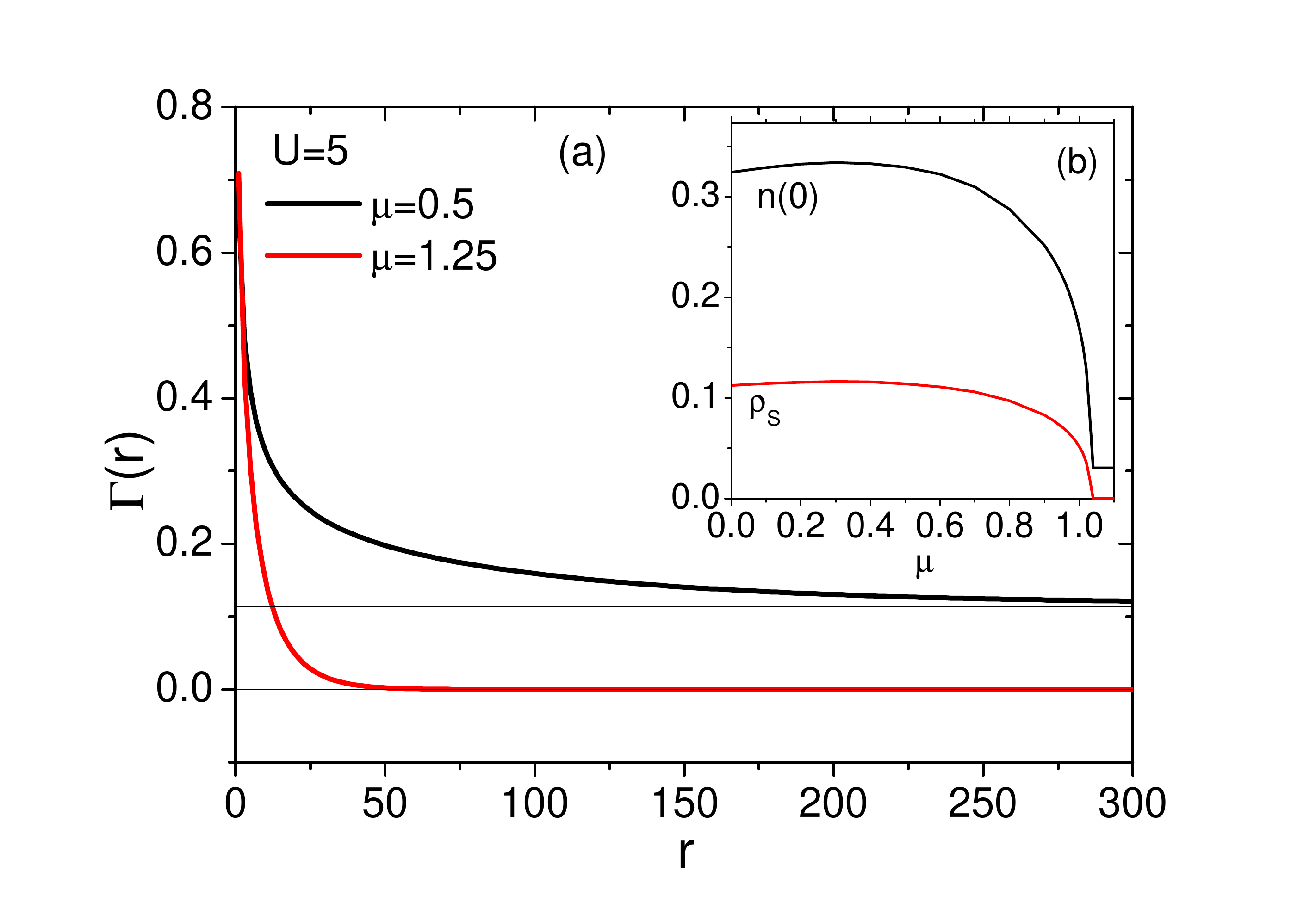}
	\caption {(Color online) (a) Decay of phase coherence correlation function $\Gamma(r)$ with respect to $r$. $\Gamma(r) \rightarrow \psi^2 $ as $r \rightarrow \infty$. (b) Variation of the condensate fraction $n(0)$ and the superfluid density $\rho_S$ across SF-MI transition. }
	\label{fig:fig5}
\end{figure}

The CMFT+DMRG method also allows us to calculate the phase coherence correlation function
\begin{equation}
\Gamma(|(j-j')|)=\frac{1}{2}(\langle a_j^\dagger a_{j'}+H.c.\rangle).
\end{equation}
Here the brackets $\langle \cdots \rangle$ denote the mean value of an observable in the system $\Psi_{GS}$. We plot $\Gamma(r=\mid (j-j')\mid )$ for the SF and the MI phases in Fig~\ref{fig:fig5}. We consider on-site interaction $U=5$, length $L=1000$ and restrict $350 \le j,j'\le 650$ so that the $j,j'$ are far from the system edges and the SF order parameters are converged in this region.  In Fig~\ref{fig:fig5}(a),  one observe that $\Gamma(r) \rightarrow \psi^2$ as $r\rightarrow \infty$. $\Gamma(r)$ decay as a power-law with $r$ in the superfluid phase and exponential in the Mott insulator phase. The Fourier transform of the phase coherence correlation function 
\begin{equation}
n(k)=\frac{1}{L^2}\sum_{j,j'}\Gamma(|j-j'|)e^{-\imath k|j-j'|}
\end{equation}
gives the number of particles of the system with a wave vector $k$. 
$n(0)$ is the condensate fraction giving the fraction of bosons occupying the superfluid ground state. We plot $n(0)$ and the superfluid density $\rho_S=\psi^2$  as a function of $\mu$ across SF-MI phase transition for $U=5$ in Fig~\ref{fig:fig5}(b). The condensate fraction increases sharply in the superfluid phase. The SF-MI transition at fixed integer density belongs to Berezinskii–Kosterlitz–Thouless transition (BKT transition)~\cite{bkt}.  The order parameter shows a discontinuity at the BKT transition. It is interesting to apply the CMFT+DMRG method to observe the discontinuity at the SF-MI transition. However, we were not successful in observing this discontinuity due to (i) the lack of known accurate relation between $\mu$ and $U$ to fix density $\rho=1$ and (ii) the need to keep a larger number of states in the DMRG procedure for closer to the transition. We could, however, observe such discontinuity across the superfluid to density wave transition in the extended Bose-Hubbard model.

\subsection{Hard-core Extended Bose-Hubbard Model}
\label{ebh-hc}
\begin{figure}[htb!]
	\centering
	\includegraphics[width=8cm, height=10cm]{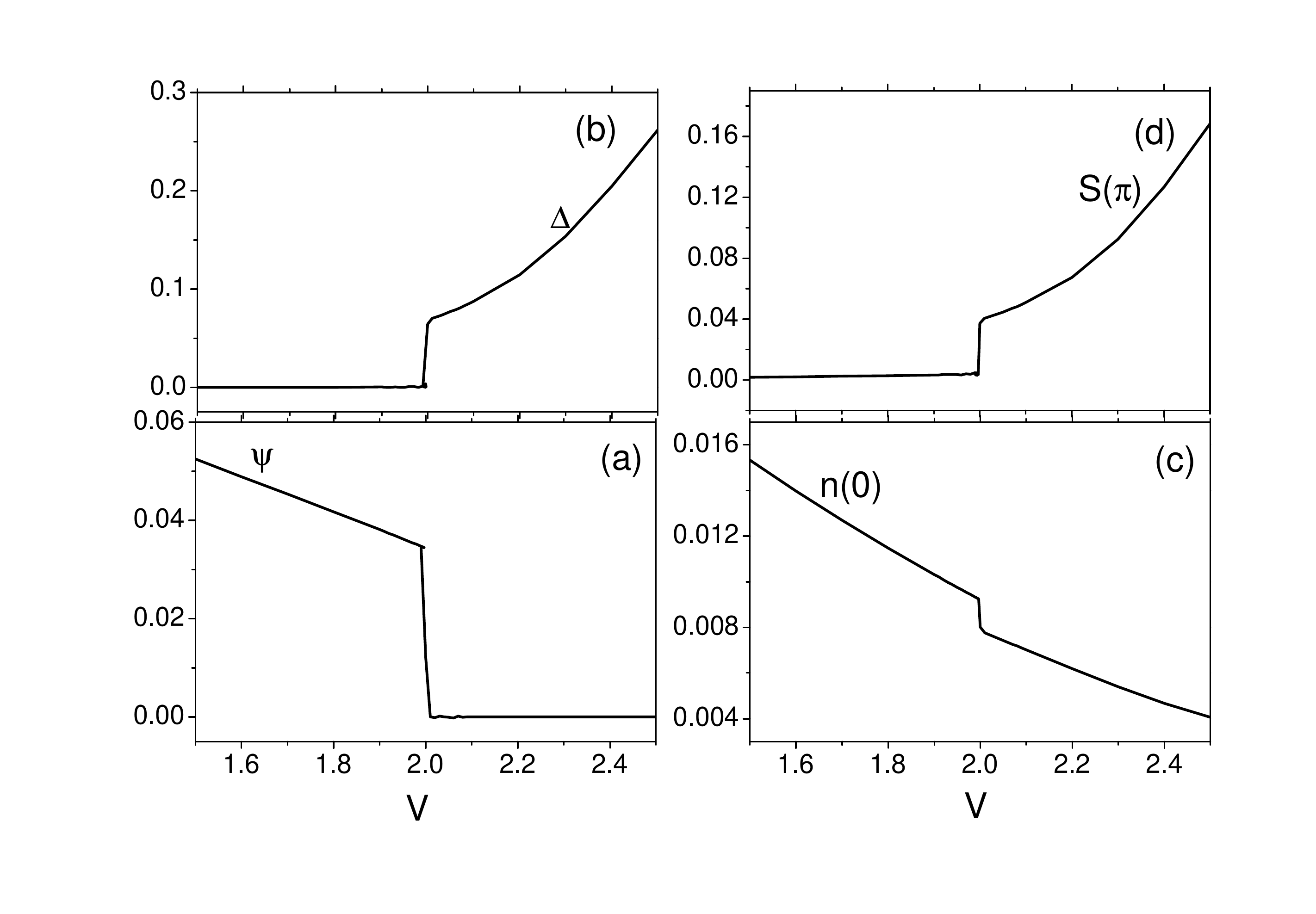}
	\caption {(a) superfluid order parameter (b) density wave order parameter (c) condensate fraction and (d) $S(\pi)$ across SF to DW phases. Decay of phase coherence correlation function $\Gamma(r)$ with respect to $r$. $\Gamma(r) \rightarrow \psi^2 $ as $r \rightarrow \infty$.}
	\label{fig:hc}
\end{figure}
The extended Bose-Hubbard model is given by
\begin{align}
\label{eq:ebh}
\hat{H} &=-t\sum_{j}\left(\hat{a}^\dagger_{j+1}\hat{a}_{j}
+\hat{a}^\dagger_{j}\hat{a}_{j+1} \right)\nonumber\\
&+\frac{U}{2}\sum_{j}\hat{n}_j(\hat{n}_j-1)+V\sum_{j}\hat{n}_j\hat{n}_{j+1}
-\mu\sum_{j}\hat{n}_j
\end{align}  
where the third term is the nearest neighbor interaction with strength $V>0$. The other terms have the same meaning as in the Eq.(\ref{eq:bh}). In the hard-core limit ($U=\infty$) and for $\mu=V$, the model (\ref{eq:ebh}) maps into spin-1/2 XXZ model, which has a BKT transition from the superfluid phase (equivalent to XY order) to the density wave phase (Ising order) at $V=2$. The density of bosons alternate between $\rho_A$ and $\rho_B\ne\rho_A$ in the density wave phase. Here $\rho_{A(B)}$ represents density of bosons at $A(B)$ sub-lattices. 

Applying the CMFT+DMRG method to the extended Bose-Hubbard model, we find working with the system having an odd number of sites is preferred over systems with an even number of lattice sites. Since the lattice has left-right symmetry, the density of bosons at the left edge site ($\rho_1$) and the right edge site ($\rho_L$) are equal. This symmetry forces $\rho_{\frac{L}{2}}=\rho_{\frac{L}{2}+1}$ if $L$ is even in all the possible phases, including the density wave phase. Since the density of bosons alternate between the nearest neighboring sites in the density wave phase, this symmetry forces density variation to have a node at the center. However, if $L$ is an odd integer, the left-right symmetry is preserved with no such restrictions.

To set up the cluster Hamiltonian for the extended Bose-Hubbard model, we decouple both the hopping and the nearest neighbor interaction terms following the procedure discussed in the earlier section. The resultant cluster Hamiltonian  is given by 
\begin{align}
\hat{H}^{C}= &=-t\sum_{j=1}^{L-1}\left(\hat{a}^\dagger_{j+1}\hat{a}_{j}
+\hat{a}^\dagger_{j}\hat{a}_{j+1} \right)\nonumber\\
&+\frac{U}{2}\sum_{j=1}^L\hat{n}_{j}(\hat{n}_{j}-1)
+V\sum_{j=1}^{L-1}\hat{n}_j\hat{n}_{j+1}
-\sum_{j=1}^{L}\mu\hat{n}_{j}\nonumber \\
&-t \left(
(\hat{a}^\dagger_{1}+\hat{a}_{1})\psi_B-\psi_A\psi_B \right)\nonumber \\ &-t\left((\hat{a}^\dagger_{L}+\hat{a}_{L})\psi_B-\psi_A\psi_B \right) \nonumber \\
&+V(\hat{n}_1+\hat{n}_L)\rho_B-\frac{V}{2}\rho_A\rho_B.
\label{eq:ebhc2}
\end{align}
where $\psi_A~(\psi_B)$ and $\rho_A~(\rho_B)$ are the superfluid order parameter and the density of bosons of  A-(B-)sub-lattices, respectively. The minimization of the ground state energy is done with respect to $\psi_{A(B)}$ and $\rho_{A(B)}$. The superfluid phase has non-zero $\psi_A$ and $\psi_B$. The density wave phase has $\psi_A=\psi_B=0$ and $\rho_A\ne\rho_B$.    

The CMFT+DMRG calculations are performed by retaining $M^S=50$ eigenstates in the left/right block reduced density matrix and taking $n_{max}=1$ ($U=\infty$). The neglected truncated weight
 $\epsilon=1-\sum_{\alpha=1}^{M^S}\omega_\alpha$ is of the order of $10^{-8}$ or less. We set the energy scale by $t=1$. 
We summarize our results in Fig.~\ref{fig:hc}. We have build the lattice starting with $L=3$ to $L=2001$ and calculate the superfluid order parameters $\psi_{A(B)}$, density wave order parameter $\Delta=|\rho_A-\rho_B|$, the condensate fraction $n(0)$ and structure factor $S(\pi)$ where 
\begin{align}
    S(\pi)=\frac{1}{L^2}\sum_{j,j'}(-1)^{|j-j'|} \langle  \hat{n_i}\hat{n}_j\rangle.
\end{align}
We find the superfluid order parameter $\psi_A=\psi_B=\psi$ is finite in the SF phase and vanishes in the density wave phase with a universal discontinuity at the transition $V=2$. Similarly, the density wave order parameter is finite in the density wave phase and vanishes in the SF phase with a discontinuity at $V=2$. The condensate fraction and $S(\pi)$ also show similar discontinuity at $V=2$. These results are consistent with the BKT transition between SF to DW phase.

\section{Conclusions}
\label{conclusions}
We have developed a novel numerical method to understand quantum phases in the one-dimensional Bose Hubbard models. This CMFT+DMRG  overcomes many limitations of the mean-field theory and the DMRG technique.  Notably, the small system size in the former and direct calculation of superfluid order parameters in the latter. The CMFT+DMRG method integrates the key features of the mean-filed theory and the DMRG method.   We apply the CMFT+DMRG method to the Bose-Hubbard model and the extended Bose-Hubbard model to test its usefulness. The Bose-Hubbard model has two phases; superfluid and Mott insulator. We identify these phases with the superfluid order parameters and the condensate fraction. Our results agree with the earlier studies done using the DMRG method. 
 The extended Bose-Hubbard model in the hard-core limit at the density of bosons equal to half shows the superfluid and the density wave phases. We identify these phases using the superfluid and the density wave order parameters, condensate fraction, and structure factor. The discontinuous jump in these physical quantities at the superfluid to density wave phase boundary confirms the BKT nature of the phase transition. This method can be extended to other models such as the soft-core extended Bose-Hubbard model, spin-1 Bose-Hubbard model, etc., to understand the exotic superfluid phases such as supersolids, polar/Ferro superfluids, and pair superfluids. The DMRG and the CMFT+DMRG methods work complementary to understand the entire phase diagram of Bose-Hubbard models. 
The former method works in the canonical ensemble and is most suitable for characterizing the gaped phases. The CMFT+DMRG, however,  works in the grand-canonical ensemble and is very useful to understand the gapless quantum phases.

\section{ACKNOWLEDGMENTS}
AD acknowledge the research grant  under the scheme for promotion of science education from DHE, Govt. of Goa.

\end{document}